\documentclass[aps,prx,showpacs,twocolumn,amsmath,amssymb,nofootinbib]{revtex4-1}
\usepackage{graphicx}
\usepackage{dcolumn}
\usepackage{bm}
\usepackage{amsmath}
\usepackage{amsfonts}
\usepackage{amssymb}
\usepackage{multirow}

\usepackage{amsthm}
\usepackage{pinlabel}
\usepackage{natbib}

\usepackage{epstopdf}
\usepackage[abs]{overpic}
\usepackage[dvipsnames]{xcolor}

\usepackage[dvipsnames]{xcolor}

\usepackage{braket}

\definecolor{specialgray}{HTML}{505050}
\definecolor{col10K}{HTML}{FFA000}
\definecolor{col300K}{HTML}{924FA4}
\definecolor{colMu}{HTML}{5278BD}
\definecolor{colMuI}{HTML}{924FA4}

\definecolor{specialgray}{HTML}{505050}
\definecolor{col10K}{HTML}{FFA000}
\definecolor{col300K}{HTML}{924FA4}
\definecolor{colMu}{HTML}{5278BD}
\definecolor{colMuI}{HTML}{924FA4}
\definecolor{newred}{HTML}{D53E4F}
\definecolor{newblue}{HTML}{5278BD}
\definecolor{newcyan}{HTML}{1EA0A0}
\definecolor{newgreen}{HTML}{5CB14E}
\definecolor{newpurple}{HTML}{924FA4}
\definecolor{newyellow}{HTML}{D1C72E}
\definecolor{neworange}{HTML}{D6923C}

\begin{document}

\title{Exploring multichannel superconductivity in ThFeAsN}
\author{Fabian Schrodi}\email{fabian.schrodi@physics.uu.se}
\author{Fairoja Cheenicode Kabeer}\email{fairoja.cheenicode-kabeer@physics.uu.se}
\author{Alex Aperis}\email{alex.aperis@physics.uu.se}
\author{Peter M. Oppeneer}\email{peter.oppeneer@physics.uu.se}
\affiliation{Department of Physics and Astronomy, Uppsala University, P.\ O.\ Box 516, SE-75120 Uppsala, Sweden}

\vskip 0.4cm
\date{\today}

\begin{abstract}
	\noindent 
	We investigate theoretically  the superconducting state of the undoped Fe-based superconductor ThFeAsN. Using input from {\em ab initio} calculations, we solve the Fermi-surface based, multichannel Eliashberg equations for Cooper-pair formation mediated by spin and charge fluctuations, and by the electron-phonon interaction (EPI). 
	Our results reveal that spin fluctuations alone, when coupling only hole-like with electron-like energy bands, can account for a critical temperature $T_c$ up to $\sim7.5\,\mathrm{K}$ with an $s_{\pm}$-wave superconducting gap symmetry, which is a comparatively low $T_c$ with respect to the experimental value $T_c^{\mathrm{exp}}=30\,\mathrm{K}$. 
	Other combinations of interaction kernels (spin, charge, electron-phonon) lead to a suppression of $T_c$ due to phase frustration of the superconducting gap. We qualitatively argue that the missing ingredient to explain the gap magnitude and $T_c$ in this material is the first-order correction to the EPI vertex. In the noninteracting state this correction 
	adopts a form supporting the $s_{\pm}$ gap symmetry, in contrast to EPI within Migdal's approximation, i.e., EPI without vertex correction, and therefore it enhances tendencies arising from spin fluctuations.
\end{abstract}

\maketitle

\section{Introduction} 

The discovery of superconductivity below $T_c^{\mathrm{exp}}\sim30\,\mathrm{K}$ in ThFeAsN\,\cite{Wang2016} has added another member to the growing family of Fe-based superconductors exhibiting large critical temperatures\,\cite{Stewart2011,Chen2014}. Most compounds in this class of materials show magnetic order in the ground state and allow for a Cooper pair condensate only upon sufficient doping or pressure\,\cite{Johnston2010,Paglione2010}. ThFeAsN takes an unusual role in this respect, since the ground state has been reported to already be non-magnetic and superconducting, without the need of external parameter tuning\,\cite{Wang2016,Shiroka2017,Albedah2017}. Additionally, it has been shown that applying pressure to this system decreases $T_c$\,\cite{Barbero2018}, which is in contrast to other Fe-based superconductors, such as bulk FeSe\,\cite{Medvedev2009}. 

Density Functional Theory (DFT) studies revealed that the electronic structure of ThFeAsN can be considered as quasi-2D, while the Fermi surface (FS) is prototypical for the Fe-based compounds\,\cite{Singh2016,Kumar2017,Sen2020_1}. It consists of hole-like bands close to the folded Brillouin zone (BZ) center, and electron-like pockets at the corners. The sign-change of the superconducting gap between electron and hole pockets is commonly associated with antiferromagnetic spin fluctuations as driving force for the Cooper pair formation\,\cite{Chubukov2008,Chubukov2012}. As for ThFeAsN, Barbero {\em et al.}\ have found a pressure independent $s_{\pm}$-wave symmetry of the order parameter by performing Knight shift and muon-spin rotation measurements\,\cite{Barbero2018}. Similarly, Wang {\em et al.}\ classified this compound as comparable to other Fe-based superconductors with respect to outcomes from magnetic susceptibility measurements, attributing an important role to antiferromagnetic spin fluctuations\,\cite{Wang2016}.

Although the discovery of ThFeAsN has triggered some interest from
both theory and experiment\,\cite{Singh2016,Kumar2017,Albedah2017,Adroja2017,Shiroka2017,Barbero2018,Li2018,Wang2018,Mao2019,Sen2020_1,Sen2020_2,Wang2020}, there has so far not been any attempt to theoretically explain superconductivity in this compound. This might be due to the apparent similarity of ThFeAsN to other Fe-based superconductors, which have been studied in great detail\,\cite{Stewart2011,Chen2014,Chubukov2012,Hirschfeld2016,Kreisel2020}. However, the aforementioned ground state properties make ThFeAsN one of very few intrinsically superconducting Fe-based compounds, along with e.g.\ LiFeAs\,\cite{Tapp2008}, which makes it worthwhile to be studied in detail. It has been found that FS nesting properties play a crucial role in the family of Fe-based superconductors\,\cite{Coldea2008,Terashima2009}, which is why we employ here a FS-based multichannel Eliashberg theory (see e.g.\ Ref.\,\cite{Bekaert2018}) to solve for the characteristics of the superconducting state. Our formalism allows to self-consistently include electron-phonon interaction (EPI), spin and charge fluctuations, all on the same footing.

We find the highest critical temperature as $T_c\simeq7.5\,\mathrm{K}$ when considering a spin-fluctuations interaction kernel, where only electron bands are coupled to hole bands. Together with an $s_{\pm}$-wave superconducting gap, exhibiting a maximum value of $1.43\,\mathrm{meV}$, this is insufficient to explain the experimental results in superconducting ThFeAsN. Taking into account the full spin-fluctuations kernel suppresses $T_c$ to $3.15\,\mathrm{K}$ due to phase frustration effects of the superconducting gap, which is a generic property of Fe-based superconductors as recently reported in Ref.\,\,\cite{Yamase2020}. In additional calculations, where we combine the influence due to spin fluctuations with EPI and charge fluctuations, no superconducting state is found for temperatures down to $2\,\mathrm{K}$. Again, the reason lies in a phase frustration of the superconducting gap. When considering the McMillan equation for electron-phonon mediated Cooper pairing, the $T_c$ vanishes in agreement with Ref.\,\cite{Sen2020_1}. Since additional Eliashberg calculations for EPI only do not lead to a finite solution either, it is apparent that the FS-based description employed here is not sufficient to explain superconductivity in this compound.

However, the ratio between characteristic phonon energy scale and minimal band shallowness suggests that vertex corrections to the EPI are important for describing the superconducting and, more generally, the interacting state. We therefore consider an expression for the renormalized vertex function due to non-adiabatic contributions to the EPI, which was derived in Ref.\,\cite{Schrodi2020_2}. By following Ref.\,\cite{SchrodiDWaveProj} for a simplified version of this function in the non-interacting state, we obtain the renormalized vertex in a `one-shot' calculation. The resulting function has, similar to the spin-fluctuation kernel, repulsive long wave vector contributions to the superconducting gap, which support the global $s_{\pm}$-wave symmetry. For small momenta, the vertex corrected EPI counteracts repulsive spin fluctuations contributions, which potentially minimizes the encountered phase frustration problems. We therefore argue that the superconducting state can potentially be explained by including spin fluctuations and vertex corrected EPI in a self-consistent Eliashberg formalism. 

The remainder of this paper is organized as follows: Starting with electronic structure calculations in Section \ref{scElEn}, we obtain two sets of energies that serve as input for the discussion on bosonic interactions in Section \ref{scInt}. The presentation of our {\em ab initio} calculations for the EPI, Section \ref{scElPh}, is followed in Section \ref{scSF} by the description of kernels due to spin and charge fluctuations. Section \ref{scSC} is subject to the discussion of the superconducting state in ThFeAsN, where we solve self-consistent multichannel Eliashberg equations in Section \ref{scEliash}. Additionally we argue why vertex corrections to the EPI are likely to contribute cooperatively with the spin-fluctuations interaction to the gap magnitude and $T_c$, Section \ref{scNonad}. Our conclusions and a brief outlook are presented in the final Section \ref{scDis}.

\section{Electronic structure calculations}\label{scElEn}

\begin{figure}[h!]
	\centering
	\includegraphics[width=1\linewidth]{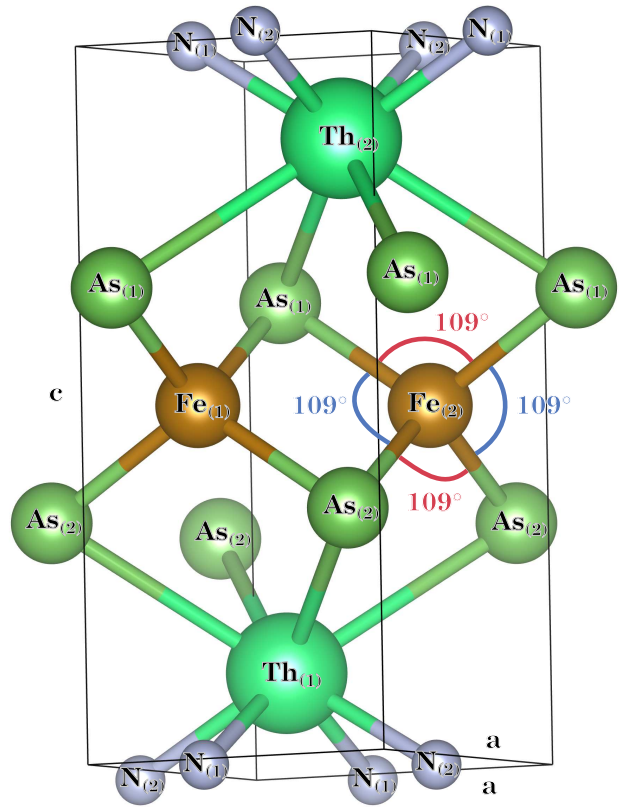}
	\caption{Two-formula unit cell of ThFeAsN as used in our DFT calculations. The angles between Fe-As bonds are highlighted in blue and red.}\label{structure}
\end{figure}

The space-group symmetry of ThFeAsN is P4/nmm, with space-group number 129, and we show the two-formula unit cell in Fig.\,\ref{structure}. The Fe atoms in this structure are in tetragonal coordination, so that neighboring Fe-As bonds enclose an angle of $109^{\circ}$. We highlight these angles in Fig.\,\ref{structure} in red and blue colors, and come back to this aspect below when discussing the distance between Fe and As planes. 

The electronic energies $\xi_{\mathbf{k},n}$ are calculated using the plane-wave DFT computational package Quantum Espresso\,\cite{qe2009,qe2017}. We choose the generalized gradient approximation with the exchange-correlation potential by Perdew, Burke, and Ernzerhod\,\cite{Perdew1996,Perdew1997}. Using scalar relativistic Projector Augmented Wave pseudopotentials with non-linear corrections for the core electrons\,\cite{Joubert1999}, we include spin-orbit coupling in our non-collinear calculation. The momentum space sampling is done on a $24\times24\times12$ Monkhorst-Pack  $\mathbf{k}$-grid and we set the kinetic energy (charge density) cutoff to $110\,\mathrm{Ry}$ ($1100\,\mathrm{Ry}$). 

\begin{figure}[h!]
	\centering
	\includegraphics[width=1\linewidth]{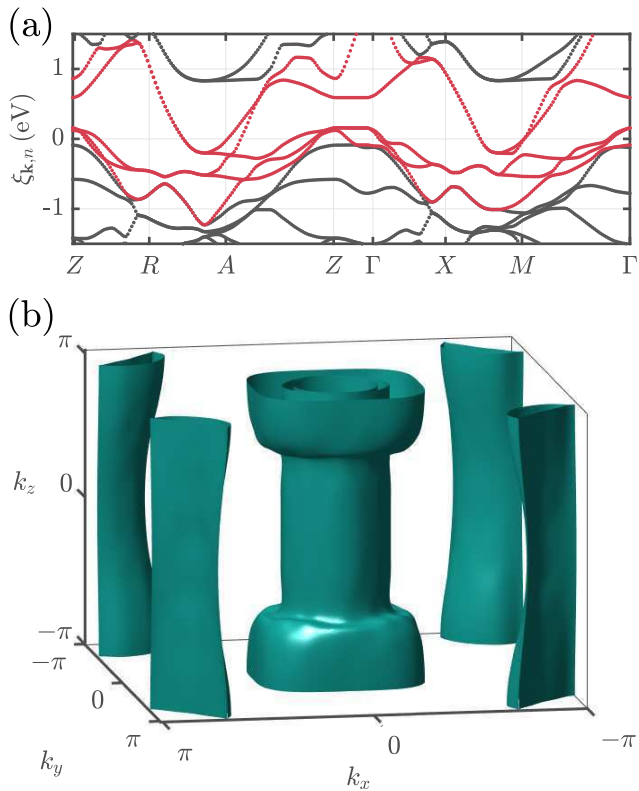}
	\caption{(a) Calculated electron dispersion of ThFeAsN with relaxed lattice parameters, shown along high-symmetry lines of the Brillouin zone, and measured relative to the Fermi energy. Electronic bands crossing the Fermi level are shown in red, the remaining bands in black. (b) Three dimensional Fermi surface corresponding to $\xi_{\mathbf{k},n}$ as shown in panel (a).}\label{elrel}
\end{figure}

As a first step we relax the structure by minimizing the forces on the atoms. Setting a convergence threshold of $10^{-8}\,\mathrm{Ry}$ for the electronic self-consistency loop, we find relaxed structural parameters $a=4.0374\,\mathrm{\AA}$ and $c=8.4307\,\mathrm{\AA}$. The corresponding electron energies are shown in Fig.\,\ref{elrel}(a) along high-symmetry lines of the BZ. The associated FS, drawn in Fig.\,\ref{elrel}(b), consists of three hole-like bands that cross the Fermi level around $Z-\Gamma$, and two electron bands at $A-X$. 
The relaxed structure is then used to calculate the lattice dynamics of the system, as we describe in Section \ref{scElPh}.

\begin{figure}[h!]
	\centering
	\includegraphics[width=1\linewidth]{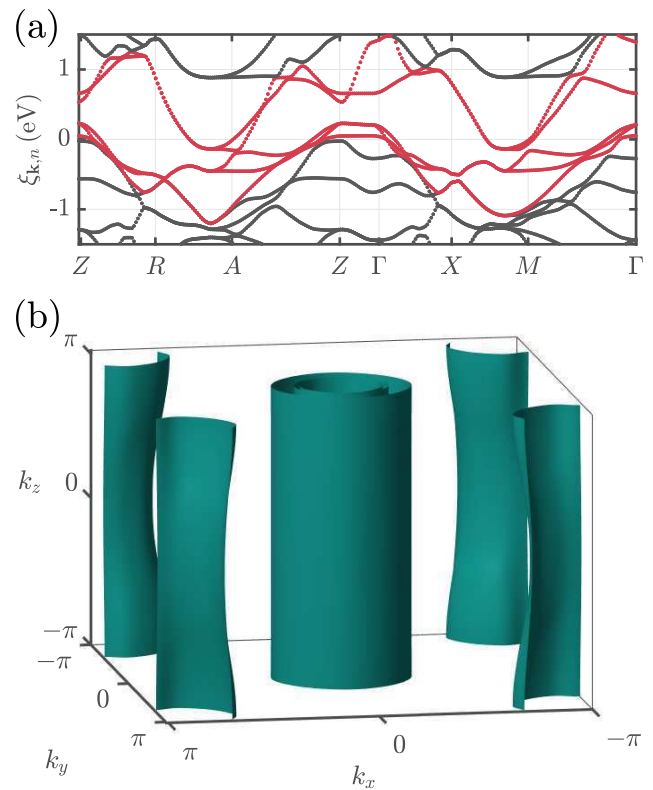}
	\caption{(a) Electron energies plotted along high-symmetry lines of the Brillouin zone, relative to the Fermi energy. The lattice parameters are fixed to the experimental values, $a=a_{\mathrm{exp}}$ and $c=c_{\mathrm{exp}}$. As in Fig.\,\ref{elrel}, we show the Fermi surface crossing bands in red and the remaining bands in black. (b) Fermi surface corresponding to panel (a).}\label{elexp}
\end{figure}

Before we discuss the ionic degrees of freedom, we note that there is a non-negligible deviation of the obtained lattice parameters when comparing to experimental findings of $a_{\mathrm{exp}}=4.0367\,\mathrm{\AA}$, $c_{\mathrm{exp}}=8.5262\,\mathrm{\AA}$\,\cite{Wang2016}. This observation in ThFeAsN was already made by other authors\,\cite{Singh2016,Sen2020_1}. We therefore perform another electronic structure calculation by fixing $a=a_{\mathrm{exp}}$ and $c=c_{\mathrm{exp}}$, the results for the band structure and FS are shown in Fig.\,\ref{elexp}.

When comparing Figs.\,\ref{elrel}(b) and Fig.\,\ref{elexp}(b) it is apparent that the relaxed structure is more dispersive along the $k_z$-direction. This stems from the fact that two energy bands in close vicinity of the Fermi level along $Z-\Gamma$ change roles, compare panel (a) in both Figures. In Fig.\,\ref{elrel}(a) we observe a dispersive band along $Z-\Gamma$ crossing the FS, and a nearly constant one slightly below. This is in contrast to a nearly flat band above the Fermi level on this momentum path, and a dispersive $\xi_{\mathbf{k},n}$ below the FS in Fig.\,\ref{elexp}(a). The energy band with higher dispersion has dominant Fe-$d_{z^2}$ orbital character, while the nearly flat band has mainly contributions from Fe-$d_{xy}$ states\,\cite{Singh2016}. As mentioned before, in the ThFeAsN compound Fe is in tetragonal coordination, hence the crystal-field splitting dictates that the $d_{z^2}$ state is lower in energy than the $d_{xy}$ state. According to this picture we conclude that Fig.\,\ref{elexp} represents results for $\xi_{\mathbf{k},n}$ that is closer to experiment. When calculating the electronic energies with relaxed lattice parameters, Fig.\,\ref{elrel}, these two bands along $Z-\Gamma$ change role, i.e. the $d_{xy}$ dominated band lies lower in energy than the $d_{z^2}$ dominated band. This behavior indicates a structural distortion in our unit cell. 

The mismatch in the unit cell height ($c=8.4307\,\mathrm{\AA}$ vs. $c_{\mathrm{exp}}=8.5262\,\mathrm{\AA}$) has the effect that the Fe-As tetragons are slightly squeezed, changing the angle between Fe-As bonds from $109^{\circ}$ to approximately $105^{\circ}$ and $119^{\circ}$ (blue and red angles in Fig.\,\ref{structure}). As a consequence, the system does not exhibit tetragonal crystal-field splitting, which is the reason that the $d_{xy}$ states fall lower in energy than the $d_{z^2}$ states. The observed mismatch between relaxed and experimental lattice structure goes in line with DFT calculations on LaFeAsO\,\cite{Mazin2008}, and has been shown to potentially influence the superconducting state significantly\,\cite{Kuroki2009}.

The results we obtain for $\xi_{\mathbf{k},n}$ (Figs.\,\ref{elrel} and \ref{elexp}) are in excellent agreement with previous works\,\cite{Singh2016,Kumar2017,Sen2020_1}. Despite the structural differences that we discussed above, we employ both, the relaxed and experimental structure in the following when looking at different superconducting channels. We use the experimental lattice parameters when being concerned with purely electronic properties, such as susceptibilities and spin-fluctuation interaction kernels, see Section \ref{scInt}. The reason for this is that such quantities require as accurate inputs as possible for yielding the correct characteristics\,\cite{Kuroki2009}. On the other hand, for electron-phonon calculations we use the relaxed structure so as to have minimized forces on each atom which lead to reliable phonon modes.

\begin{figure}[t!]
	\centering
	\includegraphics[width=1\linewidth]{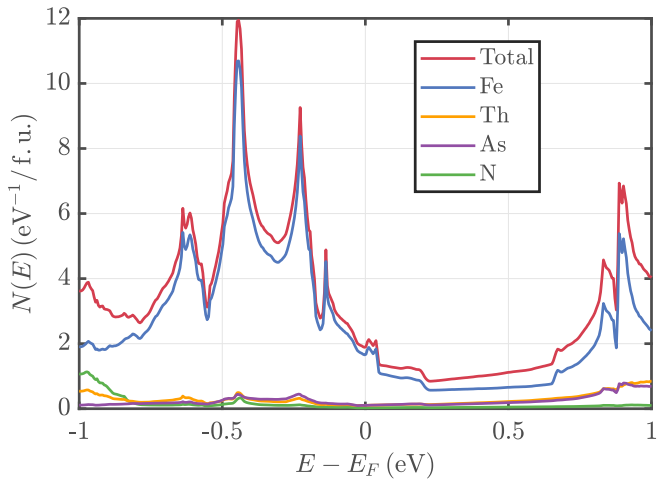}
	\caption{Atom-projected density of states per formula unit, calculated for the experimental lattice parameters (compare Fig.\,\ref{elexp}). Different colors correspond to atomic species as indicated in the legend.}\label{dos}
\end{figure}

The theory applied in later Sections \ref{scInt} and \ref{scSC} is based on electronic degrees of freedom close to the Fermi level. We therefore end the current discussion by examining the density of states (DOS) $N(E)$ in a narrow low-energy interval in Fig.\,\ref{dos}, calculated from experimental lattice parameters. The red curve, representing the full DOS per formula unit, has a characteristic shape for the family of Fe-based superconductors\,\cite{Lu2008,Nekrasov2014}, where for our undoped system the Fermi level falls into the prototypical dip of $N(E)$\,\cite{Singh2008,Singh2009}. 
As for other members of this family, most low-energy states originate from the iron atoms, shown as blue line in Fig.\,\ref{dos}. The remaining atom species play only a very minor role in the energy interval under consideration. At the Fermi level we find the values $N(0)=1.92$\,states/eV/f.u. for the total DOS, and $N_{\mathrm{Fe}}(0)=1.69$\,states/eV/f.u. for the iron DOS. These results are in reasonable agreement to related works on this material\,\cite{Singh2016,Kumar2017,Albedah2017,Sen2020_1}.

\section{Bosonic interactions} \label{scInt}

Here we discuss the EPI obtained by DFT calculations in Section \ref{scElPh} and show how to calculate spin-fluctuation kernels within the Random Phase Approximation (RPA) in Section \ref{scSF}. All calculations presented from here on, including Section \ref{scSC}, are carried out with the Uppsala Superconductivity (UppSC) code\,\cite{UppSC,Aperis2015,Schrodi2018,Schrodi2019,SchrodiMultiChan}, partially with input from the Quantum Espresso package.

\subsection{Electron-phonon interactions} \label{scElPh}

Using the relaxed lattice parameters (compare Fig.\,\ref{elrel}) we perform Density Functional Perturbation Theory calculations with Quantum Espresso. The $\mathbf{k}$-grid is chosen as in the electronic structure calculation, and we employ a $4\times4\times2$ Monkhorst-Pack $\mathbf{q}$-grid. The cutoffs for kinetic energy and charge density are fixed at $130\,\mathrm{Ry}$ and $1300\,\mathrm{Ry}$, respectively. With an effective Coulomb repulsion of $\mu^{\star}=0.136$ we find an isotropic coupling strength of $\lambda_0=0.40$. Using the modified McMillan equation \cite{McMillan1968} for the superconducting critical temperature,
\begin{align}
T_c = \frac{\omega_{\mathrm{log}}}{1.2} \mathrm{exp}\left( \frac{-1.04(1+\lambda_0)}{\lambda_0(1-0.62\mu^{\star})-\mu^{\star}} \right) \ll 1\,\mathrm{K},
\end{align}
with $\omega_{\mathrm{log}}=6.09\cdot10^{-4}\,\mathrm{Ha}$, suggests that superconductivity in this system cannot simply be explained by an isotropic, weak-coupling conventional BCS approach.

From our {\em ab inito} calculations we obtain
the branch $\nu$ resolved phonon frequencies $\omega_{\mathbf{q},\nu}$ and the couplings $\lambda_{\mathbf{q},\nu}^{(\mathrm{ep})}$. The latter are defined in terms of the electron-phonon scattering matrix elements $g_{\mathbf{q},\nu}$:
\begin{align}
\lambda_{\mathbf{q},\nu}^{(\mathrm{ep})} = 2 N(0) \frac{|g_{\mathbf{q},\nu}|^2}{\omega_{\mathbf{q},\nu}}~.
\end{align}	
By using bosonic Matsubara frequencies $q_l=2\pi T l$, $l\in\mathbb{Z}$, we can calculate the momentum and frequency dependent electron-phonon coupling as
\begin{align}
\lambda_{\mathbf{q},l}^{(\mathrm{ep})} = \sum_{\nu} \lambda_{\mathbf{q},\nu}  \frac{\omega_{\mathbf{q},\nu}^2}{\omega_{\mathbf{q},\nu}^2 + q_l^2} .
\end{align}

\begin{figure}[h!]
	\centering
	\includegraphics[width=1\linewidth]{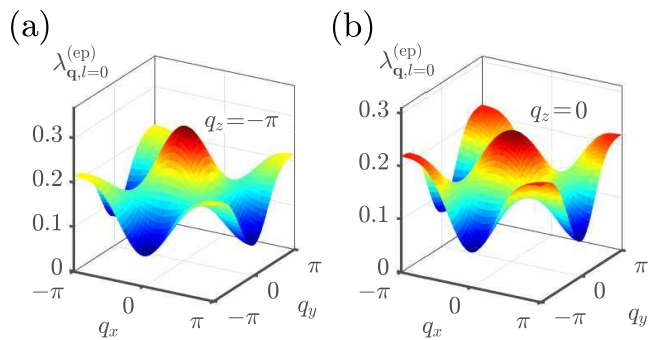}
	\caption{Zero-frequency electron-phonon coupling, calculated {\em ab initio}. (a) $q_z=-\pi$. (b) $q_z=0$.}\label{epi}
\end{figure}

In Fig.\,\ref{epi} we show $\lambda_{\mathbf{q},l=0}^{(\mathrm{ep})}=\sum_{\nu} \lambda_{\mathbf{q},\nu}^{(\mathrm{ep})}$ for $q_z=-\pi$ in panel (a), and $q_z=0$ in panel (b). In both cases we observe an enhanced magnitude at $(q_x,q_y)=(0,0)$ and $(q_x,q_y)=(-\pi,-\pi)$. The contributions at the BZ corners are comparatively smaller for $q_z=-\pi$. The overall magnitude of $\lambda^{(\mathrm{ep})}_{\mathbf{q},l=0}$ indicates a weak-coupling situation. Before using this EPI in our multichannel Eliashberg formalism in Section \ref{scSC}, we first calculate the analogue of $\lambda^{(\mathrm{ep})}_{\mathbf{q},l}$ in the spin and charge fluctuations channels.

\subsection{Spin and charge fluctuations}\label{scSF}

As mentioned before, we use from here on the electronic energies as presented in Fig.\,\ref{elexp}. Assuming that the essential physics are captured by processes at the Fermi level, we can write the band resolved bare susceptibility of the system as
\cite{Bekaert2018}
\begin{align}
\chi^{n,n'}_{\mathbf{q},l} = \sum_{\mathbf{k}} \delta(\xi_{\mathbf{k},n})\delta(\xi_{\mathbf{k}+\mathbf{q},n'}) \frac{n_F(\xi_{\mathbf{k},n}) - n_F(\xi_{\mathbf{k}+\mathbf{q},n'})}{\xi_{\mathbf{k}+\mathbf{q},n'} - \xi_{\mathbf{k},n} + iq_l} ~ \label{barechi_nn} ,
\end{align}
where we use the Fermi-Dirac function $n_F(\cdot)$ and adopt the notation $\chi^{n,n'}_{\mathbf{q},l}=\chi^{n,n'}_{\mathbf{q}}(iq_l)$. A direct approach for obtaining a band-insensitive bare susceptibility is performing a double summation as $\chi^{(0)}_{\mathbf{q},l}=\sum_{n,n'}\chi^{n,n'}_{\mathbf{q},l}$. However, it is worthwhile splitting theses summations according to certain subsets of energy bands. As encountered in Section \ref{scElEn}, there are three hole bands and two electron bands crossing the Fermi level. Let us denote the set of such bands by $h$ and $e$, respectively. We can then define
\begin{align}
&\chi^{(e-e)}_{\mathbf{q},l} = \sum_{n\in e, n'\in e} \chi^{n,n'}_{\mathbf{q},l} ,\\
&\chi^{(h-h)}_{\mathbf{q},l} = \sum_{n\in h, n'\in h} \chi^{n,n'}_{\mathbf{q},l} ,\\
&\chi^{(e-h)}_{\mathbf{q},l} = \sum_{n\in e, n'\in h} \chi^{n,n'}_{\mathbf{q},l} + \sum_{n\in h, n'\in e} \chi^{n,n'}_{\mathbf{q},l} ,\\
&\chi^{(0)}_{\mathbf{q},l} = \chi^{(e-e)}_{\mathbf{q},l} + \chi^{(h-h)}_{\mathbf{q},l} + \chi^{(e-h)}_{\mathbf{q},l} ~,
\end{align}
where the label $(e-h)$ means that all contributions involving one electron band and one hole band are included, and likewise for $(e-e)$ and $(h-h)$.

\begin{figure}[h!]
	\centering
	\includegraphics[width=1\linewidth]{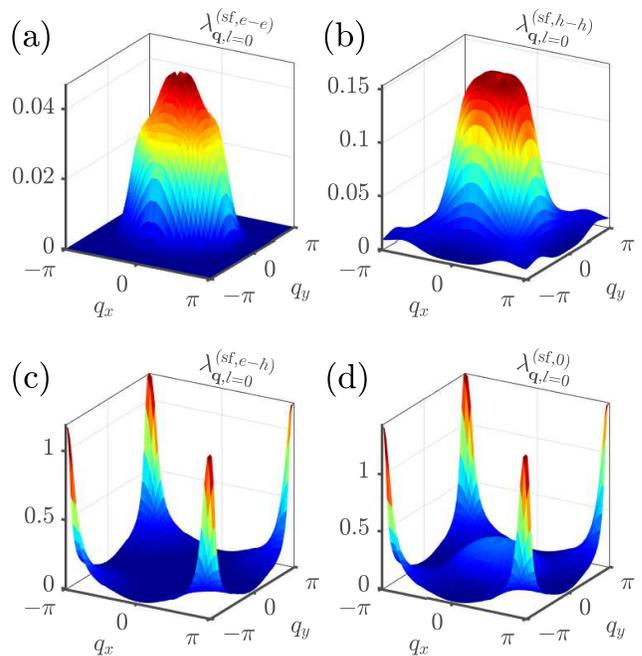}
	\caption{Spin-fluctuations kernels calculated from Eq.\,(\ref{lsf}) for $U=110\,\mathrm{meV}$ at $T=5\,\mathrm{K}$, setting $l=0$ and $q_z=0$. (a) Coupling between electron bands. (b) Coupling between hole bands. (c) Coupling between electron and hole bands. (d) Full interaction kernel due to spin fluctuations.}\label{sfkern}
\end{figure}

Next, we use the bare response of the system to calculate the interaction kernel due to spin fluctuations within the RPA under the assumption of spin-singlet Cooper pairing. Making use of the Stoner factor $U$ we compute
\begin{align}
\lambda_{\mathbf{q},l}^{(\mathrm{sf},r)} = \frac{3}{2} N(0) U^2 \frac{\chi^{(r)}_{\mathbf{q},l}}{1-U\chi^{(r)}_{\mathbf{q},l}} ~ \label{lsf} ,
\end{align}
with $r\in\{e-e,h-h,e-h,0\}$, where the magnetic instability is marked by the condition $1-U\chi^{(r)}_{\mathbf{q},l}\rightarrow0$. In a similar way the interaction kernel due to charge fluctuations is given by
\begin{align}
\lambda_{\mathbf{q},l}^{(\mathrm{cf},r)} = \frac{1}{2} N(0) U^2
\frac{\chi^{(r)}_{\mathbf{q},l}}{1+U\chi^{(r)}_{\mathbf{q},l}} . \label{lcf}
\end{align}	
We show the results for $\lambda^{(\mathrm{sf},e-e)}_{\mathbf{q},l}$ and $\lambda^{(\mathrm{sf},h-h)}_{\mathbf{q},l}$ in Fig.\,\ref{sfkern}(a) and (b), respectively, where we set $l=0$, $q_z=0$ and $U=110\,\mathrm{meV}$. Panel (c) of the same Figure shows $\lambda^{(\mathrm{sf},e-h)}_{\mathbf{q},l=0}$, i.e.\ the result for only coupling electron bands with hole bands. The full interaction kernel is drawn in Fig.\,\ref{sfkern}(d).

We observe in the first two panels of Fig.\,\ref{sfkern} that $\lambda^{(\mathrm{sf},e-e)}_{\mathbf{q},l}$ and $\lambda^{(\mathrm{sf},h-h)}_{\mathbf{q},l}$ are both peaked at $\mathbf{q}=\Gamma$. This behavior is consistent with intuition,
since the two electronic bands at the FS are located in close vicinity to each other, and likewise for the three hole bands. Therefore they are coupled via small exchange momenta, for both interband and intraband terms. Such contributions are expected to lead to a suppression of superconductivity, as noted by some of the authors in a recent work on bulk and monolayer FeSe\,\cite{Schrodi2020_3}. This `self-restraint effect' has been discussed convincingly for a more generic model system of Fe-based superconductors by Yamase and Agatsuma\,\cite{Yamase2020}, and we come back to it in Section \ref{scEliash}. Together with couplings between electron and hole bands, Fig.\,\ref{sfkern}(c), the full interaction has prominent peaks at $\mathbf{q}=M$ and a notable hump at $\Gamma$, compare Fig.\,\ref{sfkern}(d). How these features influence the superconducting solution is discussed in the following Section \ref{scSC}.

\section{Superconductivity of $\bf{ThFeAsN}$}\label{scSC}

We are now in a position to solve a self-consistent set of multichannel Eliashberg equations in Section \ref{scEliash}, where we use the interaction kernels as introduced in Section \ref{scInt}. As described in the following, the theory employed here is not capable of fully explaining the superconducting state in ThFeAsN, even though we consider the EPI, spin-fluctuations and charge-fluctuations channels. Therefore we argue in Section \ref{scNonad} that a proper inclusion of the first vertex correction to the EPI into our Eliashberg theory would likely increase the computed gap magnitude and critical temperature, so as to compare better with experimental findings.

\subsection{Eliashberg calculations}\label{scEliash}

As introduced in Section \ref{scSF}, we distinguish different contributions to the spin and charge-fluctuation interactions, according to the pair of bands they originate from. Consequently, the full kernel to be used in our Eliashberg theory similarly depends on $r\in\{e-e,h-h,e-h,0\}$. The electron-mass renormalization $Z_{\mathbf{k},m}^{(r)}$ and gap function $\Delta_{\mathbf{k},m}^{(r)}$ can be self-consistently calculated by solving
\begin{align}
Z_{\mathbf{k},m}^{(r)} &= 1 + \frac{\pi T}{\omega_m} \sum_{\mathbf{k}',m'} \frac{\delta(\xi_{\mathbf{k}'})}{N(0)}  \omega_{m'} \frac{\lambda_{\mathbf{k}-\mathbf{k}',m-m'}^{(+,r)}}{\sqrt{\omega^2_{m'} + \big(\Delta^{(r)}_{\mathbf{k}',m'}\big)^2}}, \label{zz}\\
\Delta_{\mathbf{k},m}^{(r)} &= \frac{\pi T}{Z_{\mathbf{k},m}^{(r)}} \sum_{\mathbf{k}',m'} \frac{\delta(\xi_{\mathbf{k}'})}{N(0)}  \Delta_{\mathbf{k}',m'}^{(r)} \frac{\lambda_{\mathbf{k}-\mathbf{k}',m-m'}^{(-,r)}}{\sqrt{\omega^2_{m'} + \big(\Delta^{(r)}_{\mathbf{k}',m'}\big)^2}}, \label{delta}
\end{align}
numerically. Here we use kernels $\lambda^{(\pm,r)}_{\mathbf{q},l}= \pm \lambda^{(\mathrm{sf},r)}_{\mathbf{q},l}\,(+\lambda^{(\mathrm{cf},r)}_{\mathbf{q},l})\,(+\lambda^{(\mathrm{ep})}_{\mathbf{q},l})$, where in the following we employ different combinations of spin and/or charge fluctuations, and/or EPI, as will be specified explicitly.

We already stated before that the magnetic instability of the system is marked by choosing the Stoner parameters, such that $1-U\chi^{(r)}_{\mathbf{q},l}\rightarrow0$. Therefore we can define the maximally allowed value for $U$ as
\begin{align}
U^{(r)}_{\mathrm{max}} = \big[ \mathrm{max}_{\mathbf{q},l}\,\chi^{(r)}_{\mathbf{q},l} \big]^{-1} ~.
\end{align}
For convenience we specify $U$ in the following as percentage value of its maximum, which depends on the choice of $r$, so that $U=\frac{p}{100}\cdot U^{(r)}_{\mathrm{max}}$ with $p\in(0,100)$. Further, we assume from here on a 2D system which is justified due to a very non-dispersive $k_z$-direction, compare Fig.\,\ref{elexp}. We carefully crosschecked that our results do not change qualitatively when taking into account the full 3D system.

\begin{figure}[h!]
	\centering
	\includegraphics[width=1\linewidth]{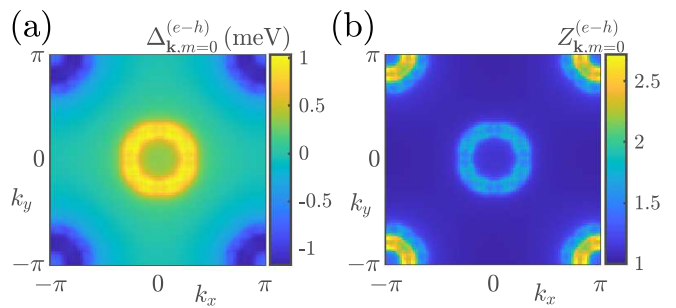}
	\caption{Self-consistent solutions to Eqs.\,(\ref{zz}) and (\ref{delta}), obtained by choosing $r=e-h$ and $p=99$, at $T=2\,\mathrm{K}$. (a) Zero-frequency superconducting gap. (b) Zero-frequency electron mass renormalization.}\label{gapsym}
\end{figure}

Let us start with the conceptually easiest case, i.e.\ considering only spin fluctuations and setting $r=e-h$. From Fig.\,\ref{sfkern}(c) we know that the interaction kernel entering in the Eliashberg equations, $\lambda^{(\pm,e-h)}_{\mathbf{q},l}= \pm \lambda^{(\mathrm{sf},e-h)}_{\mathbf{q},l}$, peaks at $\mathbf{q}=M$ and is otherwise small in magnitude and featureless. It is hence to be expected that the superconducting gap at the FS changes sign between electron and hole bands. In Fig.\,\ref{gapsym} we show our numerical solutions to Eqs.\,(\ref{zz}) and (\ref{delta}), setting $T=2\,\mathrm{K}$, $p=99$ and $r=e-h$.
Indeed, we find $\mathrm{max}_{\mathbf{k}}\,\Delta_{\mathbf{k},m=0}^{(e-h)}\sim1\,\mathrm{meV}>0$ around the BZ center in Fig.\,\ref{gapsym}(a), while a minimum negative value $\mathrm{min}_{\mathbf{k}}\,\Delta_{\mathbf{k},m=0}^{(e-h)}\sim-1.16\,\mathrm{meV}<0$ occurs around $\mathbf{k}=M$. This is the prototypical $s_{\pm}$-wave state that is shown here, which is the only stable symmetry for the superconducting gap in ThFeAsN, which is true for any choice of parameters we explored in this work.

The electron-mass enhancement is shown in Fig.\,\ref{gapsym}(b), and takes values between unity and $\sim2.7$. As apparent, the hole bands exhibit clearly smaller values of $Z_{\mathbf{k},m=0}$ than the electron bands. Consequently, we predict that electron masses at the BZ corners are noticeably higher than at the center. However, it is under question whether such a behavior can be seen in experiment, since our results change as function of $U$ and, as we discuss below, the critical temperature and maximum gap magnitude found here are lower than observed experimentally. The next step is to perform a parameter variation in $T$ and $U$ to see how these quantities affect the solutions of the Eliashberg equations.

\begin{figure}[h!]
	\centering
	\includegraphics[width=1\linewidth]{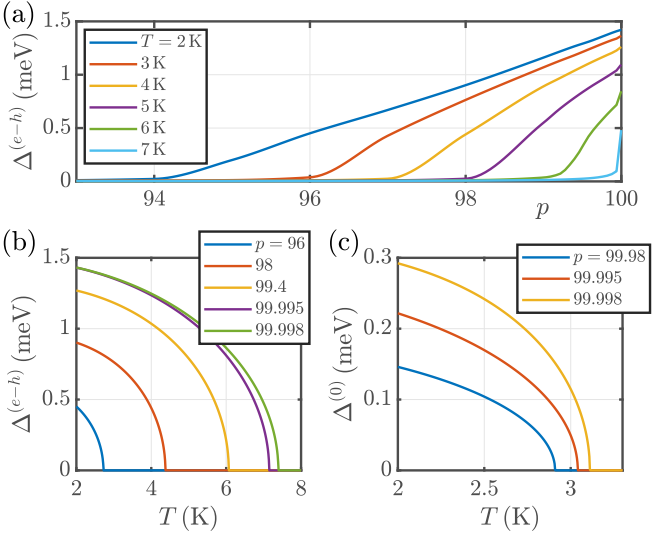}
	\caption{(a) Calculated maximum superconducting gap for $\lambda^{(\pm,e-h)}_{\mathbf{q},l}= \pm \lambda^{(\mathrm{sf},e-h)}_{\mathbf{q},l}$ as function of $p$. Different colors correspond to $T$ as indicated in the legend. (b) Same as (a), plotted against $T$ for various choices of $p$, see legend. (c) Temperature dependence of the maximum gap size, obtained for $\lambda^{(\pm,0)}_{\mathbf{q},l}= \pm \lambda^{(\mathrm{sf},0)}_{\mathbf{q},l}$, shown for three different values of $p$.}\label{gaps}
\end{figure}

In Fig.\,\ref{gaps}(a) we plot our results for $\Delta^{(e-h)}=\mathrm{max}_{\mathbf{k}}\,|\Delta_{\mathbf{k},m=0}^{(e-h)}|$, found from numerically solving Eqs.\,(\ref{zz}) and (\ref{delta}), as function of $p$ for various temperatures as indicated in the legend. It is easily observed that the maximum superconducting gap grows nearly linear with $p$, while the onset of superconductivity is strongly temperature dependent. For all solutions found in the shown parameter range, the gap function has $s_{\pm}$-wave symmetry. The reasons for a growing $\Delta^{(e-h)}$ with $p$ is that the interaction kernel is approaching a magnetic instability as $U\rightarrow U_{\mathrm{max}}$. As was shown in Ref.\,\cite{Schrodi2020_3}, the spin fluctuation kernel at the instability wave vector scales like
\begin{align}
\mathrm{max}_{\mathbf{q}}\,|\lambda^{(\pm,e-h)}_{\mathbf{q},l=0}| \propto \frac{U^2}{U_{\mathrm{max}}^{(e-h)} - U} ~, \label{lambdascale}
\end{align}
hence an increase in coupling strength is responsible for the enhancement of the gap magnitude with $p$.

It is also apparent that our theory inherits an upper limit of $\Delta^{(e-h)}$, as the gap size does not diverge for $p\rightarrow100$. The reason lies again in Eq.\,(\ref{lambdascale}): The mass renormalization scales approximately like $Z_{\mathbf{k},m=0}\propto 1+\mathrm{max}_{\mathbf{q}}\,\lambda^{(+,e-h)}_{\mathbf{q},l=0}$, while the superconducting order parameter has a similar behavior, $\phi^{(e-h)}_{\mathbf{k},m=0}\equiv \Delta^{(e-h)}_{\mathbf{k},m=0}Z^{(e-h)}_{\mathbf{k},m=0}\propto \mathrm{max}_{\mathbf{q}}\,|\lambda^{(-,e-h)}_{\mathbf{q},l=0}|$. Since the gap function is found from $\Delta^{(e-h)}_{\mathbf{k},m=0}=\phi^{(e-h)}_{\mathbf{k},m=0}/Z^{(e-h)}_{\mathbf{k},m=0}$, the divergences of both $\phi^{(e-h)}_{\mathbf{k},m=0}$ and $Z^{(e-h)}_{\mathbf{k},m=0}$ cancel approximately, and the result is only linear in the Stoner factor.

Next, let us turn to the maximum superconducting gap as function of $T$, which we show in Fig.\,\ref{gaps}(b), still for the choice $r=e-h$. Different choices for $p$ are colored as written in the legend. Note, that the values for $p$ are very critical, i.e.\ we have to consider a Stoner factor very close to $U_{\mathrm{max}}^{(e-h)}$, similar to results shown in panel (a) of the same figure. Here we find that both the maximum superconducting gap and critical temperature increase with $p$, where the largest possible values are $\Delta^{(e-h)}\sim1.43\,\mathrm{meV}$ and $T_c\sim7.5\,\mathrm{K}$. Naively one might expect that these quantities can be arbitrarily increased by choosing $U$ closer to $U^{(e-h)}_{\mathrm{max}}$, but as discussed in connection to Fig.\,\ref{gaps}(a) this is not possible due to an upper bound on the gap magnitude. The highest critical temperature found here is significantly lower than the experimental value, $T_c^{\mathrm{exp}}=30\,\mathrm{K}$\,\cite{Wang2016,Li2018}, which indicates that some important ingredients relevant for ThFeAsN are missing in our theory. 

We now turn to a different interaction kernel used for solving the Eliashberg equations, namely we include all contributions due to spin fluctuations, setting $\lambda^{(\pm,0)}_{\mathbf{q},l}=\pm \lambda_{\mathbf{q},l}^{(\mathrm{sf},0)}$. When solving Eqs.\,(\ref{zz}) and (\ref{delta}) as function of $U$ we find that a finite superconducting gap (still $s_{\pm}$-wave) can only be found if $U$ is at least $p=99.98\%$ of $U_{\mathrm{max}}^{(0)}$. The results $\Delta^{(0)}$ from three of such very critical choices are shown in Fig.\,\ref{gaps}(c) as function of $T$. It is easily observed that, compared to panel (b), both gap magnitude (max. $\sim0.3\,\mathrm{meV}$) and $T_c$ (max. $\sim3.15\,\mathrm{K}$) are reduced by more than a factor of two. This behavior is, however, not very surprising when considering the shape of the interaction kernel.

As shown earlier, the large wave vector contributions to the spin fluctuations kernel, compare Fig.\,\ref{sfkern}(d), promote an $s_{\pm}$-wave symmetry of $\Delta_{\mathbf{k},m=0}$, because they enter the equation for the superconducting gap with an overall minus sign. On the other hand, finite values of $\lambda_{\mathbf{q},l}^{(\mathrm{sf},0)}$ around the BZ center, which originate from electron-electron and hole-hole band couplings, can numerically induce a different symmetry of $\Delta_{\mathbf{k},m=0}$, e.g. a nodal $s_{\pm}$ or $d$-wave state\,\cite{Hirschfeld2016}. If the Stoner instability is not approached sufficiently close, the competition between symmetries promoted by $\lambda_{\mathbf{q}\sim\Gamma,l}^{(\mathrm{sf},0)}$ and $\lambda_{\mathbf{q}\sim M,l}^{(\mathrm{sf},0)}$ leads to a phase oscillation of the superconducting gap, which does not represent a valid solution, i.e., $\Delta=0$ meV. For more details on this self-restraint effect in Fe-based superconductors we refer to Ref.\,\cite{Yamase2020}.
	
When including EPI and/or charge fluctuations in the Eliashberg equations we do not find any superconductivity down to $T=2\,\mathrm{K}$. Choosing $\lambda_{\mathbf{q},l}^{(\pm,0)}= \lambda_{\mathbf{q},l}^{(\mathrm{ep})}$, $\lambda_{\mathbf{q},l}^{(\pm,0)}= \lambda_{\mathbf{q},l}^{(\mathrm{cf},0)}$ or $\lambda_{\mathbf{q},l}^{(\pm,0)}= \lambda_{\mathbf{q},l}^{(\mathrm{cf},0)} + \lambda_{\mathbf{q},l}^{(\mathrm{ep})}$ does not provide sufficient coupling strength to obtain a self-consistent finite superconducting gap. Taking into consideration all electronic contributions only, $\lambda_{\mathbf{q},l}^{(\pm,0)}=\pm\lambda_{\mathbf{q},l}^{(\mathrm{sf},0)} + \lambda_{\mathbf{q},l}^{(\mathrm{cf},0)}$, leads to an enhancement of the self-restraint effect because the charge-fluctuations kernel is peaked at $M$ and enters with opposite sign compared to $\lambda_{\mathbf{q},l}^{(\mathrm{sf},0)}$ into the equation for $\Delta^{(0)}_{\mathbf{k},m}$. Similarly, $\lambda_{\mathbf{q},l}^{(\pm,0)}=\pm\lambda_{\mathbf{q},l}^{(\mathrm{sf},0)} + \lambda_{\mathbf{q},l}^{(\mathrm{ep})}$ does not give a finite solution because the spin-fluctuation kernel relevant for the order parameter is decreased at both $\Gamma$ and $M$. Finally, the full multichannel interaction $\lambda_{\mathbf{q},l}^{(\pm,0)}=\pm\lambda_{\mathbf{q},l}^{(\mathrm{sf},0)} + \lambda_{\mathbf{q},l}^{(\mathrm{cf},0)} + \lambda_{\mathbf{q},l}^{(\mathrm{ep})}$ leads to $\Delta^{(0)}_{\mathbf{k},m}=0$, too, due to the aforementioned reasons. We therefore conclude that results from our self-consistent Eliashberg theory are underestimating $T_c$ and the superconducting gap magnitude. Potential cures to this behavior are discussed in the following Sections \ref{scNonad} and \ref{scDis}.

\subsection{Possibility of non-adiabatic effects}\label{scNonad}

In the previous Section \ref{scEliash} we showed that the superconducting state in ThFeAsN can neither be explained by spin fluctuations alone, nor does the inclusion of charge fluctuations and EPI lead to a complete picture. This can be explained by phase frustration (i.e., phase oscillation) of the superconducting gap, caused by repulsive spin fluctuation kernel contributions at $\Gamma$ and $M$, which both do not support the same global symmetry. We therefore want to discuss here the possibility of vertex corrections to the EPI, that potentially can contribute positively to the superconducting gap size and eventually $T_c$. In a recent work by some of the authors it was shown for a model system that unconventional gap symmetries, such as $s_{\pm}$-wave, can arise from isotropic EPI when taking vertex corrections into account\,\cite{SchrodiDWaveProj}. We therefore want to obtain here an estimate of the vertex function when considering the non-interacting state of the system.

Let us assume that the EPI is isotropic to first-order approximation, so that scattering matrix elements are given by a single number $g_0$. The characteristic energy scale of the phonon spectrum is $\omega_{\mathrm{log}}\simeq17\,\mathrm{meV}$ (value taken from DFT calculation), from which we can define the isotropic EPI kernel $V_{m-m'}=2g_0^2\omega_{\mathrm{log}}/(\omega_{\mathrm{log}}^2 + q_{m-m'}^2)$. Taking into consideration the shallowness of our system, $\epsilon_F\simeq52\,\mathrm{meV}$, we get a non-adiabaticity ratio of $\alpha=\omega_{\mathrm{log}}/\epsilon_F\sim0.33$, which is an indicator for the non-negligible relevance of vertex corrections\,\cite{SchrodiDWaveProj}. For simplicity let us further assume that the system is in the non-interacting state, which translates into an electron Green's function
\begin{align}
\big[\hat{G}^{(0)}_{\mathbf{k},n,m}\big]^{-1} = i\omega_m \hat{\rho}_0 - \xi_{\mathbf{k},n}\hat{\rho}_3 \,,
\end{align}
defined in Nambu space that is spanned by a Pauli matrix basis $\hat{\rho}_i$ ($i=0,1,2,3$). As was shown in Refs.\,\cite{Schrodi2020_2,SchrodiDWaveProj}, the electron self-energy including vertex corrections can be written as
\begin{align}
\hat{\Sigma}_{{\bf k}m}&= T\sum_{{\bf k}',n,m'}V_{m-m'} \hat{\rho}_3 \hat{G}^{(0)}_{{\bf k}',n,m'}\hat{\rho}_3 \big( 1 + g_0^2 \hat{\Gamma}^{(0)}_{\mathbf{k},\mathbf{k}',m,m'}\big) .
\end{align} 
We use here the label $(0)$ to stress that we are in a non-interacting situation. By making use of the relation $\mathbf{q}=\mathbf{k}-\mathbf{k}'$, the vertex function can be expressed as
\begin{align}
&\Gamma^{(0)}_{\mathbf{k},\mathbf{k}',m,m'} = \Gamma^{(0)}_{\mathbf{q},m,m'} \nonumber\\
&~~~ = -\frac{T}{g_0^2} \sum_{m''}V_{m'-m''}\sum_{\mathbf{k}''} \big( \gamma_{\mathbf{k}''+\mathbf{q},m''-m'+m}^{(\omega)} \nonumber\\
&~~~~~~~ \times  \gamma_{\mathbf{k}'',m''}^{(\omega)}  - \gamma_{\mathbf{k}'',m''}^{(\xi)}\gamma_{\mathbf{k}''+\mathbf{q},m''-m'+m}^{(\xi)} \big) ,
\end{align}
with the definitions $\gamma_{\mathbf{k},m}^{(\omega)}=\sum_n\omega_m/\theta_{\mathbf{k},n,m}$, $\gamma_{\mathbf{k},m}^{(\xi)}=\sum_n\xi_{\mathbf{k},n}/\theta_{\mathbf{k},n,m}$ and $\theta_{\mathbf{k},n,m}=(i\omega_m)^2-\xi_{\mathbf{k},n}^2$. We perform the Matsubara summation over index $m''$ analytically and set $m=m'$, allowing us to visualize the renormalized vertex function below.

Considering that the non-adiabatic equation for the superconducting order parameter contains the interaction kernel $-(1+g_0^2\Gamma_{\mathbf{k},\mathbf{k}',m,m'})$\,\cite{SchrodiDWaveProj}, we want to find a rough estimate of the scattering strength $g_0$ by following the approach of Ref.\,\cite{Adroja2017}. According to McMillan, the electron-phonon coupling constant can to first order approximation be written as
\begin{align}
\lambda = \frac{1.04 + \mu^{\star}\log(\Theta_D/1.45T_c)}{(1-0.62\mu^{\star}) \log(\Theta_D/1.45T_c) - 1.04} \,,
\end{align}
where $\Theta_D$ is the Debye temperature. Inserting $\mu^{\star}=0.136$, $\Theta_D=332\,\mathrm{K}$\,\cite{Albedah2017} and $T_c=T_c^{\mathrm{exp}}=30\,\mathrm{K}$\,\cite{Wang2016} gives $\lambda\simeq1.6$. From here we can solve for the scattering strength via $g_0=\sqrt{\lambda\omega_{\mathrm{log}}/2N(0)}$.

\begin{figure}[t!]
	\centering
	\includegraphics[width=0.8\linewidth]{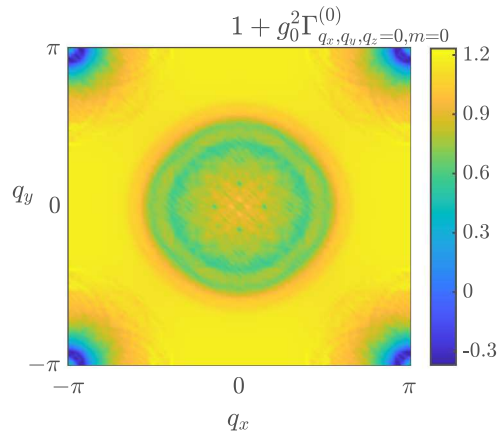}
	\caption{Estimate for the renormalized electron-phonon vertex $1+g_0^2\Gamma^{(0)}_{\mathbf{q},m}$ at $m=0$ and $q_z=0$, computed for $\omega_{\mathrm{log}}=16.7\,\mathrm{meV}$ at $T=10\,\mathrm{K}$.}\label{epivertex}
\end{figure}

We show the two dimensional zero-frequency result for $(1+g_0^2\Gamma_{\mathbf{q},m=0})^{(0)}$ in Fig.\,\ref{epivertex}, obtained at $T=10\,\mathrm{K}<T_c^{\mathrm{exp}}$.
At $\mathbf{q}=M$ the vertex correction is negative, while in the remaining BZ our result is strictly positive. These contributions enter the equation for the order parameter in a repulsive and attractive way, respectively. Therefore, in combination with the spin-fluctuations kernel, see Section \ref{scSF}, we get an enhanced repulsive interaction at $\mathbf{q}=M$, which supports the $s_{\pm}$-wave symmetry of the gap. On the other hand, the repulsive small-$\mathbf{q}$ contribution of Fig.\,\ref{sfkern}(d) might likely be compensated 
by an attractive contribution from $\Gamma_{\mathbf{q},0}^{(0)}$ around the BZ center, such that the `self-restraint' effect is minimized\,\cite{Yamase2020}.

\section{Discussion}\label{scDis}

In summary, a self-consistent FS-based Eliashberg theory, including spin and charge fluctuations, and EPI, can not explain the critical temperature in ThFeAsN as it is observed experimentally. In Section \ref{scEliash} we show that this is due to phase frustration in the superconducting gap due to competing contributions to the interaction kernel. From the strong similarity to other Fe-based superconductors, especially concerning FS properties, we conclude that such frustration behavior is likely to be generic for this family of compounds. However, even in the most simplified case of solely including spin-fluctuation couplings between electron and hole bands, where no phase frustration occurs, we find a maximum $T_c=7.5\,\mathrm{K}$, which is still significantly smaller than $T_c^{\mathrm{max}}=30\,\mathrm{K}$.

As mentioned above, our Eliashberg calculations lead to an underestimation of both $T_c$ and superconducting gap magnitude, due to a phase frustration in the gap, caused by competing short and long range wave vector contributions to the spin-fluctuation interaction. It is possible that an inclusion of orbital dependent matrix elements in the calculation of bare susceptibilities would improve this aspect significantly\,\cite{Graser2009,Heil2014}. Such matrix elements can alter the momentum structure of the susceptibility, and therefore also of the charge and spin kernels, such that the problematic features at the BZ center might be suppressed. However, at this point we can only speculate about the influence of orbital content on the level of susceptibilities, because the self-restraint effect in Fe-based superconductors was only demonstrated in calculations without such matrix elements\,\cite{Yamase2020}.

Another potentially important aspect is the level on which the RPA formalism is introduced. Here we use a global Stoner parameter, i.e.\ only a single quantity to tune in order to find a valid solution for the superconducting state. More general Hubbard Hamiltonians can include additional physics, such as Hundness and intra-/inter-orbital hopping, as e.g.\ employed in Refs.\,\cite{Takimoto2002,Takimoto2004,Kubo2007,Kemper2010}. Not only can such modeling provide more insights into the important physics of the system, but, additionally, a larger phase space for parameter exploration is likely to allow the correct solution of the interacting state. For a realistic anisotropic, full bandwidth and multi-band/-orbital description one might employ the formalism of Ref.\,\cite{Schrodi2020_3}. However, a faithful tight-binding fit to the electronic dispersion is crucial for this approach, which is why we did not attempt it here. 

The potential boost for $T_c$ due to vertex corrected EPI as discussed in Section \ref{scNonad} has its root in the quasi-2D character of the electronic energies, together with a degree of non-adiabaticity, that suggests a treatment of EPI beyond Migdals approximation as important. For representative model systems of high-$T_c$ materials, including Fe-based superconductors, it has been shown that a well-nested FS leads to self-consistent unconventional superconductivity, mediated by isotropic electron-phonon scattering\,\cite{SchrodiDWaveProj}. Although only a fully self-consistent calculation in ThFeAsN can test our conjecture, the renormalized vertex function obtained here resembles to a good degree recent model systems\,\cite{SchrodiDWaveProj}.
We are hence led to conclude that non-adiabatic corrections are likely effective to enhance spin-fluctuations mediated Cooper pairing in ThFeAsN, while supporting the $s_{\pm}$ symmetry of the gap.

\begin{acknowledgments}
	F.\,S. thanks Raquel Esteban-Puyuelo for valuable advice on interpreting the results of our {\em ab initio} calculations. This work has been supported by the Swedish Research Council (VR), the R{\"o}ntgen-{\AA}ngstr{\"o}m Cluster, and the Knut and Alice Wallenberg Foundation (Grant No.\ 2015.0060). The calculations were enabled by resources provided by the Swedish National Infrastructure for Computing (SNIC) at NSC Link\"oping, partially funded by the Swedish Research Council through grant agreement No.\ 2018-05973.	
\end{acknowledgments}

\bibliographystyle{apsrev4-1}

%

\end{document}